\documentclass[11pt, letter]{article}
\usepackage{jheppub}
\usepackage{amsmath, amsfonts, amsthm, amssymb, graphicx}
\usepackage{color, xcolor}
\usepackage{comment}
\usepackage[utf8]{inputenc}
\usepackage{lipsum}
\usepackage{centernot}
\usepackage{ulem}
\usepackage{mathtools}
\usepackage{enumitem}
\def\be{\begin{equation}}
\def\ee{\end{equation}}
\def\ben{\begin{equation*}}
\def\een{\end{equation*}}

\def\del{\partial}

\newcommand{\citeseq}{\cite{Kaloper:2013zca, Kaloper:2014dqa, Kaloper:2014fca, Kaloper:2015jra, Kaloper:2016yfa, Kaloper:2016jsd, DAmico:2017ngr, Niedermann:2017cel, Coltman:2019mql} }

\DeclareMathAlphabet\mathbfcal{OMS}{cmsy}{b}{n}
\renewcommand{\arraystretch}{1.5}
\newcommand\figref{Fig.~\ref}

\newcommand\secref{Sec.~\ref}

\newcommand\tabref{Tab.~\ref}
\renewcommand{\eqref}[1]{Eq.~(\ref{#1})}

\newcommand{\xip}{\relax\ifmmode \xi_\phi \else $\xi_\phi$ \fi}
\newcommand{\sa}{\relax\ifmmode \mathfrak{a} \else $\mathfrak{a}$ \fi}
\renewcommand{\sb}{\relax\ifmmode \mathfrak{b} \else $\mathfrak{b}$ \fi}
\renewcommand{\sc}{\relax\ifmmode \mathfrak{c} \else $\mathfrak{c}$ \fi}
\newcommand{\nobs}{\relax\ifmmode n_{\rm obs} \else $n_{\rm obs}$\fi}
\title{CMB constraints on monodromy inflation at strong coupling}
\author[1]{Edmund J. Copeland, }
\author[1]{Francesc Cunillera, }
\author[1]{Adam Moss,}
\author[1]{Antonio Padilla}    

\affiliation[1]{School of Physics and Astronomy, 
University of Nottingham, Nottingham NG7 2RD, UK} 

\emailAdd{ed.copeland@nottingham.ac.uk}
\emailAdd{francesc.cunilleragarcia@nottingham.ac.uk}
\emailAdd{adam.moss@nottingham.ac.uk}
\emailAdd{antonio.padilla@nottingham.ac.uk}

\abstract{We carry out a thorough numerical examination of field theory monodromy inflation at strong coupling.  We perform an MCMC analysis using a Gaussian likelihood, fitting multiparameter models using CMB constraints on the spectral index and the tensor to scalar ratio. We show that models with uniquely positive Wilson coefficients are ruled out. If there are  coefficients that can take on both signs, there can be a cancellation of terms that flattens the potentials and allows one to satisfy current data, and forecasts with strong constraints on the tensor to scalar ratio.  Models of field theory monodromy are naturally enhanced to include a mechanism for canceling off radiative corrections to vacuum energy, via vacuum energy sequestering (VES).  Although they include a much larger parameter space, we find that a similar numerical examination yields  no significant change in the Bayesian evidence for VES enhanced models, with naturalness considerations making them  more attractive from a theoretical perspective. 
}

\begin{document}
\maketitle

\section{Introduction}
Inflation provides the most successful explanation for the isotropy and homogeneity observed in the cosmic microwave background (CMB) to date. Indeed, modern observations of the CMB \cite{1807.06211} are in agreement with an early epoch of accelerated expansion driven by a slowly rolling scalar field. On top of this, the quantum fluctuations during this inflationary epoch lead to large-scale structure formation which is consistent with the structure we observe in the CMB and in galaxy distributions (for nice reviews see \cite{Baumann:2009ds} and \cite{Weinberg:2008zzc}).

Single field inflationary models are quite remarkable in their simplicity. For a canonical scalar the observational consequences of these models are determined by the first and second derivative of the potential. Of particular interest are large field inflation models which provide some of the simplest and most calculable models of the early universe \cite{Linde:1983gd, Freese:1990rb}. Due to the super-Planckian excursions, these models are very sensitive to CMB measurements. On the other hand, protecting the effective description from quantum corrections is non-trivial due to the large values of the inflaton field. A possible approach is to use symmetries that protect the scalar field from UV effects. One such approach within string theory is monodromy inflation \cite{0803.3085, 0808.0706}. In these scenarios the inflaton is an axion, a pseudoscalar with a discrete shift symmetry allowing one to explore large field values without losing control of the effective description. 

A field theory description of axion monodromy was first introduced in \cite{hep-th/0410286, hep-th/0507215} and its implications  for inflation later explored in \cite{arXiv:0810.5346, arXiv:0811.1989, arXiv:1101.0026}. The  effects of higher order corrections and strong coupling were investigated in detail in \cite{1607.06105} through the Kaloper-Lawrence (KL) Lagrangian (cf. \eqref{eq:L_EFT}) allowing constraints based on effective monomial potentials  to be derived in \cite{1709.07014}. It should be noted that although it is a consistent EFT, it remains unclear whether or not there exists a  consistent string embedding of the KL model \cite{1611.00394}.

In this paper, we begin by generalising the analysis of \cite{1709.07014} to derive the observational implications of the KL models. In particular, we do not limit our analysis to effective monomials but rather consider the full Lagrangian in \eqref{eq:L_EFT} and show that one can consistently truncate its sums to polynomials of some degree. The exact degree of the polynomials is fixed by demanding that the error associated to the truncation of the infinite sums, {\it i.e.} the higher order operators that we neglect, is smaller than the observational error margins of the data we use to fit the observables, in our case the \textit{Planck} 2018 data of \cite{1807.06211}. This leads to an effective description of KL in terms of degree 20 polynomials and the ability to discuss the predictions of a large class of these models in full generality. We will limit ourselves, however, to two-derivative theories, meaning that the relevant cosmological observables that limit the parameter space of the theory will be: the spectral index $n_s$, the tensor-to-scalar ratio $r$ and the amplitude of scalar perturbations $\mathcal{A}_s$. It would be interesting to consider the introduction of higher order derivatives operators  since it would add further bounds to the parameter space coming from, say, non-gaussianities and the speed of propagation of scalar perturbations. This is left for future work. 

We also extend our study in another important direction, allowing us to probe a proposed mechanism \citeseq     for solving the cosmological constant problem \cite{Padilla:2015aaa}.
This follows from the realisation that monodromy inflation models could be deformed in a natural way to  give rise to vacuum energy sequestering (VES)  \cite{1806.04740}. 
Within the VES framework, radiative corrections  to vacuum energy are reabsorbed through new rigid degrees of freedom. The theory remains locally indistinguishable from General Relativity, while global dynamics are modified. The mechanism itself, initially inspired by Tseyltin's model of duality symmetric strings \cite{Tseytlin:1990hn}, 
is reminiscent of so-called decapitation, at least in a field theory context \cite{hep-th/0209226, 1706.04778}. 

In \cite{1806.04740}, it was shown that the KL model could be made compatible with VES by introducing a second monodromy into the theory. The EFT would then consist of a heavy monodromy, whose scalar would play the role of a constant dilaton enforcing the global VES constraints on the system, and a light sector, which would contain the inflaton. Then, the cancellation of the vacuum energy loops can be accomplished via the sequestering mechanism at low energies, consistent with a particular structure for the inflationary sector. The recipe for this EFT description requires multiple axions, together with their respective discrete shift symmetries, and a hierarchy between a heavy and a light sector. These ingredients appear naturally in string theory, which prompts the interesting question of whether a similar mechanism would be available to string axion monodromy models of inflation.  One of the main goals of this paper is to investigate the impact of VES deformations on KL models of monodromy inflation, via their imprint on the CMB. 

\begin{figure}[ht!]
\centering
\includegraphics[width=0.7\textwidth]{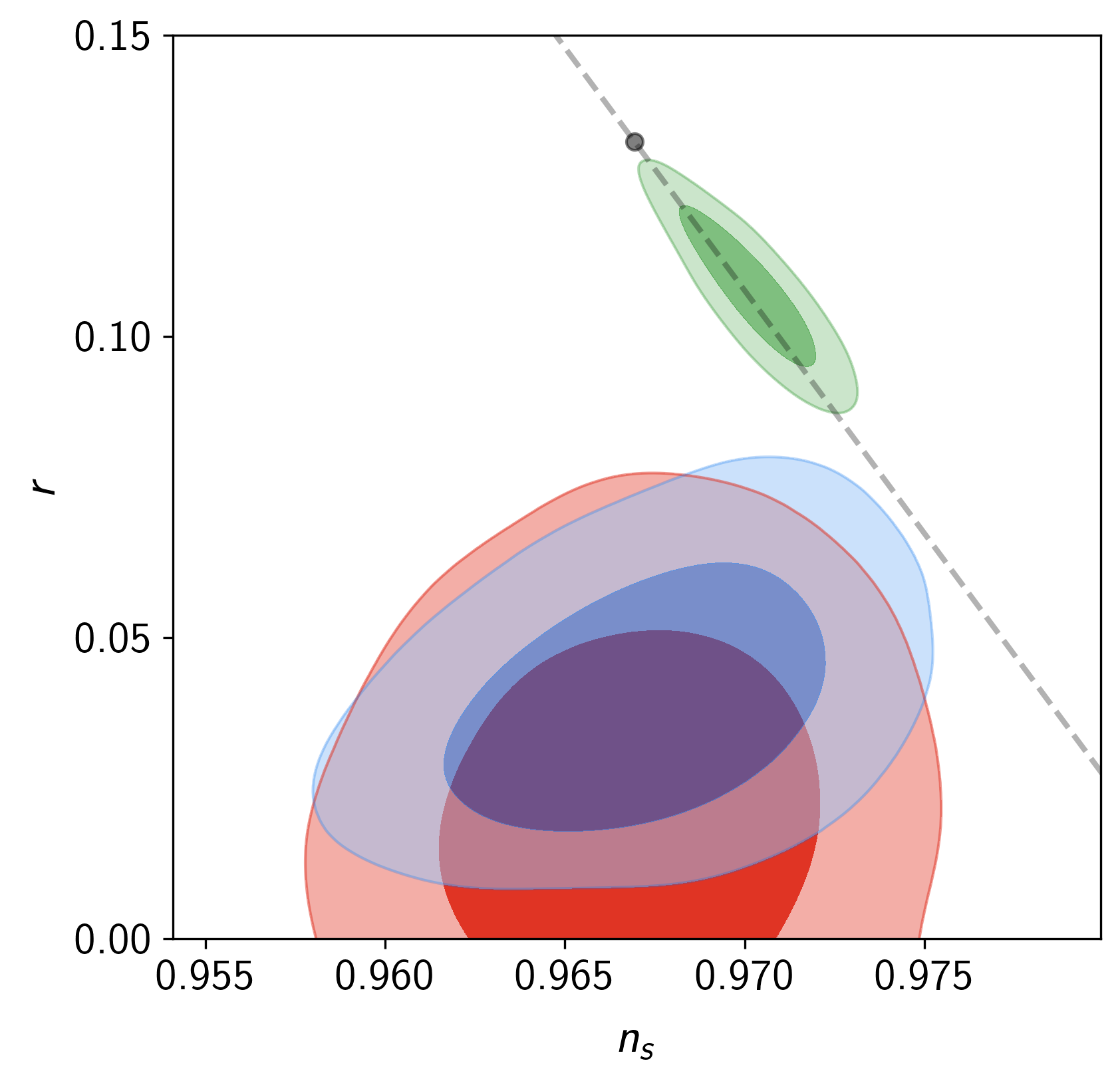}
\caption{\textit{Planck} data (red) along with KL models: with only positive coefficients (green) and with positive-negative coefficients (blue). Superimposed is the monomial line, {\it i.e.} the $n_s$-$r$ predictions for potentials of the form $V\sim \phi^p$, with $p=2$ and $N_{\star}=60$ given by the filled circle.}
\label{fig:intro_kl}
\end{figure}

Overall, after the numerical analysis, we find that KL models both with and without VES deformations,  are able to fit current \textit{Planck} data and forecasted CMB Stage 3 and Stage 4 pseudo-data. We remark that this is in contrast with the effective monomial results of \cite{1709.07014} which would be disfavoured at a $2\sigma$ level by current observations. The key difference is in the more general form of the Lagrangian allowed by our analysis. Critically, fitting the \textit{Planck} data requires both positive and negative coefficients in \eqref{eq:L_EFT}. The monomial potentials of \cite{1709.07014} can be mapped to our analysis with a single positive coefficient in the potential, all other terms being turned off.  As can be seen in \figref{fig:intro_kl}, the monomial line lies outside of the data and crosses the region with only positive coefficients. Allowing for negative coefficients provides a much more efficient way of lowering $r$ while keeping $n_s$ on the observational range. We are also able to show that the introduction of VES is not in tension with observations. Therefore, VES enchanced KL models have the added benefit of including a mechanism to solve the cosmological constant problem in a natural way without any observational drawbacks.

The structure of the paper is as follows: in \secref{sec:KL} we introduce the KL models, then in  \secref{sec:seq_KL} we follow \cite{1806.04740} in introducing a second monodromy and present the form of the KL Lagrangian with VES enhancement. In \secref{sec:non_can_inf} we present the relevant equations for discussing slow roll monodromy inflation at strong coupling, with and without VES enhancements.  In \secref{sec:caveats} and \secref{sec:MCMC} we describe our numerical approach in some detail and present the results for the predictions of the KL models, with and without a VES sector, for current \textit{Planck} data and forecasted CMB Stage 3 and Stage 4 pseudo-data. Finally, we conclude in \secref{sec:conc}.
\section{The Kaloper-Lawrence model \label{sec:KL}}
In \cite{1607.06105, 1101.0026, 1709.07014}, the authors build upon earlier work \cite{hep-th/0410286, hep-th/0507215} to construct an inflation model inspired by axion flux monodromy \cite{0803.3085, 0808.0706}. The starting point for the KL model is a massive $U(1)$ gauge theory given by
\be
\mathcal{L}=-\frac{1}{48}F^2-\frac{m^2}{12}\left(A_{\mu\nu\alpha}-h_{\mu\nu\alpha}\right)^2+\frac{m}{6}\phi \frac{\epsilon^{\mu\nu\alpha\beta} }{\sqrt{-g}} \partial_\mu h_{\nu\alpha\beta}\ , 
\ee
where $F_{\mu\nu\alpha\beta}=4 \del_{[\mu} A_{\nu\alpha\beta]}$ is the four-form field strength, $\epsilon^{\mu\nu\alpha\beta}$ is the totally antisymmetric Levi-Civita symbol defined such that $\epsilon^{0123}=1$, and $\phi$ is an axion with periodicity $\phi \to \phi+2\pi f$, with $f$ left to be determined. Indices are raised and lowered with respect to the metric $g_{\mu\nu}$. The equation of motion for $\phi$ yields $h=db$, with $b_{\mu\nu}$ the Stueckelberg field invariant under $b\rightarrow b+d\lambda$, and $h$ is it's three form field strength. Integrating out the Stueckelberg field through its equation of motion yields
\be
\mathcal{L}=-\frac{1}{48}F^2-\frac{1}{2}\left(\partial \phi\right)^2 + \frac{m}{24}\frac{\epsilon^{\mu\nu\alpha\beta} }{\sqrt{-g}}\phi F_{\mu\nu\alpha\beta}\ . \label{eq:L_U1}
\ee
We now introduce a Lagrange multiplier $Q$, fixing $F_{\mu\nu\alpha\beta}=4 \del_{[\mu} A_{\nu\alpha\beta]}$ on shell via a term $\int d^4 x \frac{1}{24} Q \epsilon^{\mu\nu\alpha\beta} (F_{\mu\nu\alpha\beta} -4 \del_{[\mu} A_{\nu\alpha\beta]})$. As the four-form now enters the Lagrangian algebraically, it can be integrated out,  with $Q$ now identified with its magnetic dual. This allows us to rewrite the theory in terms of a pair of pseudo-scalars,
\be
\mathcal{L}=-\frac{1}{2}(\partial\phi)^2-\frac{m^2}{2}\left(\phi+\frac{Q}{m}\right)^2-\frac{Q}{6}\frac{\epsilon^{\mu\nu\alpha\beta} }{\sqrt{-g}}\partial_\mu A_{\nu\alpha\beta}\ , \label{eq:L_KS}
\ee
where $Q$ is constrained to be constant on-shell. In the presence of membranes, it is quantised in units of the membrane charge $q$, $\langle Q\rangle=2\pi N q$ for integers $N$. The Lagrangian is invariant under a discrete shift symmetry
\be
\phi\ \rightarrow\ \phi+2\pi f\ , \qquad Q\ \rightarrow \ Q- 2\pi q\ , \label{eq:sym}
\ee
where $f=q/m$. 

The theory \eqref{eq:L_KS} describes quadratic inflation, which would be in tension with current observational bounds \cite{1807.06211}. One can ease the tension with observations by exploiting corrections due to higher order operators. Completing the theory, by writing all the possible higher order terms allowed by symmetries, one finds \cite{1607.06105}
\begin{align}
\mathcal{L}_{KL}&=-\frac{1}{2}(\partial\phi)^2-\frac{m^2}{2}\left(\phi+\frac{Q}{m}\right)^2-\sum_{k>1}\frac{a_k}{2^k\, k!}\left(\frac{4\pi}{M^2}\right)^{2k-2}\left(\partial\phi\right)^{2k}\notag \\
&-\sum_{\substack{n>2\\}}\frac{b_n}{n!}\left(\frac{4\pi}{M^2}\right)^{n-2}\left(m\phi+Q\right)^n -\sum_{\substack{k,n>1\\}}\frac{c_{k,n}}{2^k\, k!n!}\left(\frac{4\pi}{M^2}\right)^{2k+n-2}\left(m\phi+Q\right)^n \left(\partial\phi\right)^{2k}\ \notag  \\
&-\frac{Q}{6}\frac{\epsilon^{\mu\nu\alpha\beta} }{\sqrt{-g}}\partial_\mu A_{\nu\alpha\beta}\ , \label{eq:L_EFT}
\end{align}
where all the coefficients are taken to be $\mathcal{O}(1)$ by naturalness and $M$ is the EFT cut-off. By using naive dimensional analysis (NDA) \cite{Manohar:1983md, 1601.07551} one can find the appropriate factors of $4\pi$ that guarantee that the higher order corrections remain under control as long as we probe energies below the cut-off.

Crucially,  the theory permits a small window  $M/\sqrt{4\pi}<m\phi + Q< M^2$ where it is strongly coupled (lower bound)  but the effective description  remains under control (upper bound). In this strongly coupled regime, the corrections become important and can flatten the overall potential. This argument was used in \cite{1709.07014} to motivate a description for the potential as a shallower-than-quadratic monomial, {\it i.e.} $V\sim \varphi^p$ with $p<2$ where $\varphi$ is the canonical inflaton. A priori, monomial inflationary potentials with $p<2$ tend to push the models to higher values of the spectral index $n_s$ putting them in tension with data. As we will show, the results of \cite{1709.07014} are disfavoured by \textit{Planck} 2018. Nonetheless, one might hope that the more general character of \eqref{eq:L_EFT} might capture regimes that are not trivially given by the monomial potentials, significantly altering their location in the $n_s-r$ plane, as we will show in the later sections. 
\section{The Kaloper-Lawrence model with VES enhancement \label{sec:seq_KL}}
The monodromy inflation models of the previous section can be deformed in a very natural way, giving rise to an emergent mechanism for solving the cosmological constant problem \cite{1806.04740}.   The mechanism for cancelling off radiative corrections to the vacuum energy is achieved via vacuum energy sequestering \citeseq. In this section, we shall review how field theory monodromy can be deformed to give rise to VES and how this yields the desired cancellation of radiative corrections to the cosmological constant.   VES enhanced monodromies can also be thought of as an extension of VES  to high energies, perhaps pointing the way to an embedding of VES in fundamental theory.  

We start with the leading order KL theory  \eqref{eq:L_U1} and introduce a small symmetry breaking deformation parametrised by a spurion term
\be
\mathcal{L}=-\frac{1}{48}F^2-\frac{1}{2}\left(\partial \phi\right)^2 + \frac{m}{24}\frac{\epsilon^{\mu\nu\alpha\beta} }{\sqrt{-g}}\phi F_{\mu\nu\alpha\beta}-\frac{1}{2}\bar{m}^2\phi^2\ , \label{eq:L_KSd}
\ee
where $\bar{m}$ is the mass for the spurion and $\left(\frac{\bar{m}}{m}\right)\ll1$. Such a term may  arise from integrating out loops of heavy matter if there exists  a suitable coupling between this and  the inflaton. In any event, one can now dualise the deformed Lagrangian and apply the NDA scheme to find 
\begin{align}
\hspace{-4em}\mathcal{L}&=-\sum_{k\geq1}\frac{a_k}{2^k\, k!}\left(\frac{4\pi}{M^2}\right)^{2k-2}\left(\partial\phi\right)^{2k}-\sum_{\substack{n\geq2\\}}\frac{b_n}{n!}\left(\frac{4\pi}{M^2}\right)^{n-2}\left(m\phi+Q\right)^n\notag \\
&-\sum_{\substack{k,n\geq1\\}}\frac{c_{k,n}}{2^k\, k!n!}\left(\frac{4\pi}{M^2}\right)^{2k+n-2}\left(m\phi+Q\right)^n \left(\partial\phi\right)^{2k}-\sum_{l\geq2}\frac{d_l}{l!}\left(\frac{4\pi}{M^2}\right)^{l-2}\left(\bar{m}\phi\right)^l\notag \\
&-\sum_{k,l\geq1}\frac{\sa_{k,l}}{2^k\, k!l!}\left(\frac{4\pi}{M^2}\right)^{2k+l-2}\left(\bar{m}\phi\right)^l \left(\partial\phi\right)^{2k}-\sum_{\substack{l\geq1, \\n\geq2}}\frac{\sb_{n,l}}{l!n!}\left(\frac{4\pi}{M^2}\right)^{l+n-2} \left(\bar{m}\phi\right)^l \left(m\phi+Q\right)^n\notag \\
&-\sum_{\substack{k,n,l\geq1\\}}\frac{\sc_{k,n,l}}{2^n\,k!l!n!}\left(\frac{4\pi}{M^2}\right)^{2k+l+n-2}\left(\bar{m}\phi\right)^l \left(m\phi+Q\right)^n \left(\partial\phi\right)^{2k} -\frac{Q}{6}\frac{\epsilon^{\mu\nu\alpha\beta} }{\sqrt{-g}}\partial_\mu A_{\nu\alpha\beta}\  \ , \label{eq:L_KL_seq}
\end{align}
 and the labels for the coefficients comply with
\begin{align}
&2k+l+n\geq3\ . \notag
\end{align}
The Lagrangian contains a pseudo symmetry: it will remain invariant under shifts of the axion field and the magnetic dual  as in \eqref{eq:sym} as long as we transform the spurion
\be
 \bar{m}\rightarrow\bar{m}-\frac{2\pi q}{\phi}\frac{\bar{m}}{m}\ .
\ee
To simplify the notation henceforth, it will be convenient to define $\mu \equiv {M\over \sqrt{4\pi}}$ and the following dimensionless quantities
\begin{align}
\kappa \equiv \frac{\bar{m}}{m}\ ,  \qquad \xi \equiv \frac{m\varphi}{\mu^2} \ , \qquad \xip \equiv \frac{m\phi}{\mu^2} \ ,
\end{align}
where $\varphi \equiv \phi+Q/m$ will be our inflaton field. If we now perform a derivative expansion and truncate the Lagrangian at second order, the EFT becomes
\begin{align}
\mathcal{L} &=-\frac12 \mathcal{Z}_{\text{eff}}\left(\xi,\kappa\,\xip\right)(\del \varphi)^2-\mu^4\mathcal{V}_{\text{eff}}\left(\xi,\kappa\,\xip\right) -\frac{Q}{6}\frac{\epsilon^{\mu\nu\alpha\beta} }{\sqrt{-g}}\partial_\mu A_{\nu\alpha\beta}\ 
\end{align}
where the explicit forms of the dimensionless wavefunction renormalisation and the potential are respectively given by
\begin{gather}
\mathcal{Z}_{\text{eff}} \equiv 1+\sum_{n\geq1}\frac{c_{1,n}}{n!}\xi^{n} +\frac{\sa_{1,n}}{n!}\left(\kappa\,\xip\right)^{n} + \sum_{l,n\geq1}\frac{\sc_{1,n,l}}{l!n!}\xi^{n}\left(\kappa\,\xip\right)^l \ , \\
\mathcal{V}_{\text{eff}} \equiv \sum_{n\geq2}\frac{b_n}{n!}\xi^{n}+\frac{d_n}{n!}\left(\kappa\,\xip\right)^n+\sum_{\substack{n\geq2\\l\geq1}}\frac{\sb_{n,l}}{l!n!}\xi^{n}\left(\kappa\, \xip\right)^l\ ,
\end{gather}
with $b_2=1$. 

To make contact with VES,  we introduce a second monodromy where the field $\hat{\phi}$ is given a very heavy mass $\hat{m}$ 
\be
\hat{ \mathcal{L}}=\frac{M_g^2}{2}R -\frac{1}{48}\hat{F}_{\mu\nu\alpha\beta}^2-\frac{1}{2}(\partial\hat{\phi})^2+\frac{\hat{m}}{4!}\hat{\phi}\epsilon^{\mu\nu\alpha\beta}\hat{F}_{\mu\nu\alpha\beta}+\frac{1}{2}\hat{g}^2R\hat{\phi}^2\ ,\qquad \label{eq:L_KSgd}
\ee
where $\hat{g}$ is a dimensionless coupling constant of the axion with gravity, which we expect to be $\hat{g}\leq1$ for energies below the UV cutoff. We will also demand that $\hat{m}\gg M$, thus for the range of validity of the EFT the field $\hat{\phi}$ is forced to lie at the minimum of its potential. This implies that once we find the dual theory to this Lagrangian in much the same way as we did for \eqref{eq:L_KSd}, the gravitational contribution goes as 
\be
\hat{\mathcal{L}}=\frac{M_g^2}{2}\left[1+\left(\frac{\hat{Q}}{\hat{m}\hat{M}}\right)^2\right] R-\frac{\hat Q}{6}\frac{\epsilon^{\mu\nu\alpha\beta} }{\sqrt{-g}}\partial_\mu \hat A_{\nu\alpha\beta}\ ,
\ee
where we have defined the energy scale $\hat{M}=\frac{M_g}{\hat{g}}$. As promised, from the viewpoint of the EFT, adding the second monodromy is equivalent to minimally coupling the inflaton to gravity. The full action for the theory is now given by  \cite{1806.04740}
\be
S=S_{\text{eff}}+S_g+S_F+S_m\ ,
\ee
where
\begin{align}
&S_{\text{eff}}=\int d^4x \sqrt{-g}\left[-\frac{1}{2}\mathcal{Z}_{\text{eff}}\left(\xi,\kappa\,\xip\right)\left(\partial\varphi\right)^2-\mu^4\mathcal{V}_{\text{eff}}\left(\xi,\kappa\,\xip\right)\right]\ , \\
&S_g=\int d^4x \sqrt{-g}\ \frac{M_g^2}{2}\left[1+\left(\frac{\hat{Q}}{\hat{m}\hat{M}}\right)^2\right] R\ , \\
&S_F=-\int d^4x \frac{\epsilon^{\mu\nu\alpha\beta}}{6}\left[Q\partial_\mu A_{\nu\alpha\beta}+\hat Q\partial_\mu \hat A_{\nu\alpha\beta}\right]\ .
\end{align}
 and $S_m$ is the action for all additional matter fields, including the Standard Model, minimally coupled to the metric $g_{\mu\nu}$.  To understand how vacuum energy is cancelled,  consider the path integral, and in particular the integration over the threes forms and the Lagrange multipliers, 
$$Z=\int \ldots [ \mathcal{D} Q ] [\mathcal{D} \hat Q] [\mathcal{D} A] [\mathcal{D} \hat A]  e^{i (S_{\text{eff}}+S_g+S_F+S_m)}.$$
Integrating out the three forms suppresses all local variations of the Lagrange multipliers,  $\partial_\mu Q=\partial_\mu \hat{Q}=0$, such that the path integral is reduced to 
$$Z=\int \ldots [ d Q ] [d \hat Q]   e^{i (S_{\text{eff}}+S_g+S_m-Qc-\hat Q \hat c) }.$$
where the Lagrange multipliers  are fixed to be spacetime constants, although we still allow  for their global variation in the path integral, and $c=\int F$, $\hat c=\int \hat F$ correspond to the global values of the flux (see Appendix A of  \cite{1903.07612} for more details).The equations of motion include two local equations coming from the variation of the metric, $g_{\mu\nu}$, and the inflaton, $\varphi$,
\begin{align}
&M_g^2\left[1+\left(\frac{\hat{Q}}{\hat{m}\hat{M}}\right)\right]^2G_{\mu\nu}=T_{\mu\nu}+T^{\varphi}_{\mu\nu}\ , \label{eq:Einstein_VES} \\
&\nabla_\mu\left[\mathcal{Z}_\text{eff}\nabla^\mu\varphi\right]- \frac{m}{2\mu^2} (\mathcal{Z}_{\text{eff}, 1}+\kappa \mathcal{Z}_{\text{eff}, 2})(\del \varphi)^2-m \mu^2 (\mathcal{V}_{\text{eff}, 1}+\kappa \mathcal{V}_{\text{eff}, 2})=0 
\end{align}
where $\mathcal{Z}_{\text{eff}, i}$ denotes the partial derivative of  $\mathcal{Z}_{\text{eff}}$ with respect to its {\it i-th} argument (similarly for $\mathcal{V}_{\text{eff}}$). The energy momentum tensors for the matter fields and the inflaton are given respectively by $T_{\mu\nu}$  and 
$$T^{\varphi}_{\mu\nu}=\mathcal{Z}_\text{eff}\del_\mu\varphi \del_\nu \varphi +p_\varphi g_{\mu\nu}$$
where $p_\varphi=- \frac12  \mathcal{Z}_\text{eff}(\del \varphi)^2-\mu^4\mathcal{V}_{\text{eff}}$ is the inflaton pressure. In addition to these local equations of motion, there are two global constraints coming from the global variation of the Lagrange multipliers, giving
\begin{align}
c&=\frac{\del S_\text{eff}}{\del Q}=-\frac{\kappa}{\mu^2}\int d^4x \sqrt{-g}\left[-\frac{1}{2}\mathcal{Z}_{\text{eff},2}\left(\partial\varphi\right)^2-\mu^4\mathcal{V}_{\text{eff},2}\right] ,\label{var_Q1}\\ 
\hat c&=\frac{\del S_\text{g}}{\del \hat Q}=\int d^4 x \sqrt{-g} R\left(\frac{M_g}{\hat{M}}\right)^2\frac{\hat{Q}}{\hat{m}^2} \ ,\label{var_Q2}
\end{align}
Defining the normalised spacetime average $\langle .\rangle$ as
\be
\langle Y\rangle\equiv \frac{\int d^4x \sqrt{-g}\ Y}{\int d^4x \sqrt{-g}}\ ,
\ee
and taking ratios of  \eqref{var_Q1} and \eqref{var_Q2} we arrive at the following constraint on the long wavelength mode of the Ricci scalar
\be
\langle R\rangle=\frac{\kappa}{\hat Q} \left(\frac{ \hat{m}\hat{M}}{\mu M_g}\right)^2\left\langle \frac{1}{2}\mathcal{Z}_{\text{eff},2}\left(\partial\varphi\right)^2+\mu^4\mathcal{V}_{\text{eff},2} \right \rangle  \frac{\hat c }{c} \equiv R_\infty(c, \hat c) \label{eq:VES_constraint}\ .
\ee
This constraint ensures that the large scale scalar curvature is controlled by the flux terms and is not  affected by large radiative corrections to the vacuum energy. To see the implications of this, we take traces and spacetime expectation values of the field equations, to arrive at the VES version of the Einstein equations
\be
\kappa_g^2 G_{\mu\nu}=T_{\mu\nu}-\frac14 \langle T \rangle g_{\mu\nu}+\mathcal{Z}_\text{eff} \partial_\mu\varphi\partial_\nu\varphi-g_{\mu\nu}\left(\delta\lambda +\Delta \Lambda_{\text{eff}}\right)\ . \label {VESEE}
\ee
Here $\kappa_g^2 \equiv M_g^2\left(1+\left(\frac{\hat{Q}}{\hat{m}\hat{M}}\right)^2\right)$  is the effective gravitational coupling, $\delta \lambda=\langle p_\varphi \rangle-p_\varphi$ measures local fluctuations of  the inflation pressure and
\be
\Delta \Lambda=\frac14 \left[ \kappa_g^2 R_\infty(c, \hat c)+\langle \mathcal{Z}_\text{eff} (\del \varphi)^2 \rangle \right]
\ee
is a global cosmological constant term that depends on the global flux, but is independent of the vacuum energy.  To see how this equation is independent of the Standard Model vacuum energy, we decompose the energy-momentum tensor of matter into a  vacuum energy piece and local excitations (likes stars and planets) {\it i.e.} $T_{\mu\nu}=-V_{\text{vac}}g_{\mu\nu}+\tau_{\mu\nu}$.  One can easily verify that $V_\text{vac}$ drops out of the VES Einstein equations \eqref{VESEE}, giving 
\be
\kappa_g^2 G_{\mu\nu}=\tau_{\mu\nu}-\Lambda_\text{eff} g_{\mu\nu}+\mathcal{Z}_\text{eff} \partial_\mu\varphi\partial_\nu\varphi-\delta \lambda g_{\mu\nu}\ . \label {VESEE2}
\ee
where the residual cosmological constant $\Lambda_\text{eff}=\Delta \Lambda+\frac14 \langle \tau \rangle$. Large radiative corrections to the vacuum energy  arising from matter loops do not induce large corrections to $\Lambda_\text{eff}$. Note that the VES mechanism does not explain why the residual cosmological constant is small but it does explain why it is radiatively stable. The scenario is now reminiscent of chiral symmetry and fermion masses in effective field theories - this explains why radiative corrections to the mass are under control but does not explain or predict the value of the mass itself.

\section{Equations for slow-roll inflation at strong coupling  \label{sec:non_can_inf}}
From the previous sections, we see that the dynamics of strongly coupled monodromy inflation, with and without VES enhancement, is captured by an effective Einstein equation of the form
\be
\kappa_g^2 G_{\mu\nu}=-\Lambda_\star g_{\mu\nu}+\mathcal{Z}_\text{eff} \partial_\mu\varphi\partial_\nu\varphi- g_{\mu\nu}\left[ \frac12  \mathcal{Z}_\text{eff}(\del \varphi)^2+\mu^4\mathcal{V}_{\text{eff}}\right]\ . \label {EE}
\ee
and an  inflaton equation
\be
\nabla_\mu\left[\mathcal{Z}_\text{eff}\nabla^\mu\varphi\right]- \frac12 \mathcal{Z}_{\text{eff}, \varphi}(\del \varphi)^2-\mu^4\mathcal{V}_{\text{eff}, \varphi}=0 
\ee
The original Kaloper-Lawrence model, coupled to General Relativity, is obtained by taking the $\kappa \to 0$ limit for $\mathcal{Z}_\text{eff}$ and $\mathcal{V}_\text{eff}$, and {\it tuning} $\Lambda_\star$ to vanish. Radiative corrections to the vacuum energy would destablise the latter, which is, of course, the statement of the cosmolgical constant problem  \cite{Padilla:2015aaa}.  In terms of our numerical analysis,  the relevant Wilson coefficients in the two potentials are
$
\{b_n,c_n\}
$.
In contrast, when we include VES enhancement,  the cosmological constant is given by
\be
\Lambda_\star=\frac14 \kappa_g^2 R_\infty(c, \hat c)-\frac14 \langle \mathcal{Z}_\text{eff} (\del \phi)^2+4 \mu^4 \mathcal{V}_\text{eff} \rangle 
\ee
Once again we fix $\Lambda_\star$ to vanish, although this result now remains stable under radiative corrections to the vacuum energy thanks to the VES mechanism. The potentials also admit corrections from the VES deformations, parametrised by the combination
\be
\kappa\, \xip \equiv  \frac{m\kappa}{\mu^2}\left(\varphi-\frac{Q}{m}\right)=\frac{\bar m}{\mu^2} \phi\ , \notag
\ee
which comes from the fact that the spurion term is related to the axion field $\phi$ rather than the inflaton field $\varphi$.  Since the inflaton is dominated by the flux contribution, the VES deformation terms are heavily suppressed with respect to the KL terms, {\it i.e.} $\bar{m}\phi \ll m\varphi$. One can then approximate $\mathcal{Z}_{\rm eff} \simeq \mathcal{Z}_{\rm eff}(\xi,0)$ and $\mathcal{V}_{\rm eff} \simeq \mathcal{V}_{\rm eff}(\xi,0)$, neglecting the VES contributions altogether. In contrast, derivatives of these quantities {\it will} see a non-trivial contribution from the VES  corrections. For example, under the previous considerations the effective potential becomes
\be
\mathcal{V}_{\text{eff}}=\sum_{n\geq2}\left(b_n + \sb_{n,1} \kappa\, \xip\right)\frac{\xi^n}{n!} + \mathcal{O}(\kappa^2 \xip^2) \simeq \mathcal{V}_{\text{eff}}(\xi,0)\ ,
\ee
since $\kappa\,\xip\ll 1$. However, its first derivative receives a non-trivial contribution, at leading order in $\kappa$, from the $\sb_{n,1}$ term
\be
{\partial \over \partial \varphi} \mathcal{V}_{\text{eff}}= {m\over \mu^2}\sum_{n\geq2}\left(n\,b_n + \sb_{n,1}\kappa\, \xi\right)\frac{\xi^{n-1}}{n!} + \mathcal{O}(\kappa \xip) \ , \label{eq:d_V_eff}
\ee
which is only suppressed by $\kappa$. In general, the {\it p-th} derivative of the potential will obtain a contribution of the form (for $p\geq 2$)
\be
{\partial^{p} \over \partial \varphi^p} \mathcal{V}_{\text{eff}}= \left({m\over \mu^2}\right)^p\sum_{n\geq p}\left[(n-p+1)\,b_n + p\, \sb_{n,1}\kappa\, \xi\right]\frac{\xi^{n-p}}{(n-p+1)!} + \mathcal{O}(\kappa \xip) \ .
\ee
A similar discussion applies to derivatives of $\mathcal{Z}_{\rm eff}$. Overall, we find that the relevant Wilson coefficients for our numerical analysis are
\be
\{b_n, \sb_n, c_n, \sc_n\}\ 
\ee
where we have set $\sb_n\equiv \sb_{n,1}$, $c_{1,n}\equiv c_n$ and $\sc_{1,n,1}\equiv \sc_n$  to avoid cluttering the notation in the remainder of this paper. The VES enhancement is contained in the $\sb_n$ and $\sc_n$.

For general values of the coefficients in the effective Lagrangian of \eqref{eq:L_KL_seq}, there is no closed form for the integral defining the canonical inflaton field
\be
\psi \equiv \int\sqrt{\mathcal{Z}_{\text{eff}}}\; d\varphi =\frac{\mu^2}{m}\int\sqrt{1+\sum_{n\geq1}\frac{c_{n}}{n!}\xi^{n}}\; d\xi \ . \label{can_field}
\ee
For this reason it is convenient to study the dynamics of  slow-roll inflation in a non-canonical frame. Let us now briefly review the relevant formulae, 
beginning with  the geometrical slow-roll parameter 
\begin{align}
\epsilon \equiv -\frac{\dot{H}}{H^2}=\frac{\dot{\varphi}}{2H}\left[\frac{\frac{m}{2\mu^2}\mathcal{Z}'_{\text{eff}}\dot{\varphi}^2+\mu^2 m\ \mathcal{V}'_{\text{eff}}+\mathcal{Z}_{\text{eff}}\ddot{\varphi}}{3H^2}\right]\simeq\frac{1}{2}\frac{1}{\mathcal{Z}_{\text{eff}}}\left(\frac{\mathcal{V}'_{\text{eff}}}{\mathcal{V}_{\text{eff}}}\right)^2\ , \label{eq:non_can_eps}
\end{align}
where prime denotes $\del_\phi$. In the last step we have used the slow roll approximation for the non-canonical field $\varphi$
\begin{align}
&\hspace{2.5em}\frac{1}{2}\mathcal{Z}_{\text{eff}}\dot{\varphi}^2\ll\mu^4\left(\mathcal{V}_{\text{eff}}\right)\ ,\notag \\
&\frac{m}{2\mu^2}\mathcal{Z}'_{\text{eff}}\left(\dot{\varphi}\right)^2+\mathcal{Z}_{\text{eff}}\ddot{\varphi}\ll\mu^2 m\mathcal{V}'_{\text{eff}}\ ,
\end{align}
Similarly, one can derive the form of the slow-roll parameter $\eta$ from its definition as
\be
\eta \equiv \frac{\dot{\epsilon}}{H \epsilon}\simeq \frac{M_{pl}^2\, m^2}{\mu^4}\frac{1}{\mathcal{Z}_{\text{eff}}}\left(\frac{\mathcal{V}_{\text{eff}}''}{\mathcal{V}_{\text{eff}}}-\frac{1}{2}\frac{\mathcal{V}_{\text{eff}}'\mathcal{Z}'_{\text{eff}}}{\mathcal{V}_{\text{eff}}\mathcal{Z}_{\text{eff}}}\right)\ . \label{eq:non_can_eta}
\ee
It is now straightforward to express the spectral index $n_s$, the scalar-to-tensor ratio $r$ and the amplitude of scalar perturbations $\mathcal{A}_s$ using the usual formulae
\begin{gather}
n_s=1-6\epsilon+2\eta\ , \label{eq:noncan_ns} \\
r=16\epsilon\ ,\label{eq:noncan_r}\\
\mathcal{A}_s=\frac{\mu^4}{24\pi^2 M_{pl}^4}\frac{\mathcal{V}_{\text{eff}}}{\epsilon}\left(1+\frac{2}{3}\epsilon+\mathcal{O}(\epsilon^2)\right)\ .
\label{eq:scalar_pert}
\end{gather}
where the form of the slow-roll parameters is taken as in \eqref{eq:non_can_eps} and \eqref{eq:non_can_eta} .

Finally, the number of e-foldings is given by
\be
dN \equiv  \frac{1}{\epsilon H}dH \underset{\text{SR}}{\simeq}d\psi\frac{V_\text{eff}}{V_{\text{eff},\psi}}\ ,
\ee
where the final expression comes from assuming slow roll. This yields an expression
\be \label{eqn:efolds}
N_{\star}=-\frac{\mu^4}{M_{pl}^2 m^2}\int_{\psi(\xi=\xi_{\star})}^{\psi(\xi=\xi_e)}d\xi\ \mathcal{Z}_{\text{eff}} \frac{\mathcal{V}_\text{eff}}{\mathcal{V}_{\text{eff},\xi}}\ .
\ee
computed in the non-canonical frame. Here $\xi_e$ is the value of the field at the end of inflation and $\xi_{\star}$ its value when the pivot scale, $k_\star =0.05 \text{ Mpc}^{-1}$ crosses the horizon.
\section{Numerical strategy and caveats \label{sec:caveats}}
The most immediate obstacle to developing a sound numerical strategy in the strongly coupled regime is the infinite tower of Wilson coefficients that appear in the relevant potentials. Indeed, for the KL model, both the wave renormalisation factor, $\mathcal{Z}_\text{eff}$, and the effective potential, $\mathcal{V}_\text{eff}$, can be schematically reduced to an infinite series of the form
\be
\mathcal{C}_\text{eff}(\xi)=\sum_{n}\frac{c_n}{n!}\xi^n\ . \label{eq:char_ser}
\ee
VES enhanced models also contain an additional series of the same schematic form, given in terms of $\xi_\phi$ as opposed to $\xi$. Of course, series of the form \eqref{eq:char_ser}  converge for $c_n\sim\mathcal{O}(1)$. However, for arbitrary coefficients, they  cannot be expressed in closed form in terms of elementary functions. To get around this,   we  truncate the potentials at some finite order $n_\text{obs}$, replacing $\mathcal{C}_\text{eff}(\xi)$ with 
\be
\bar{\mathcal{C}}_{{\rm eff}}(\xi)=\sum_{n\leq \nobs}\frac{c_{n}}{n!}\xi^n
\ee
and compute the corresponding observables $\bar n_s$ ands $\bar r$.  Such a truncation will inevitably induce an error in the potentials, given schematically as 
\be
\Delta \mathcal{C}_{{\rm eff}}=\sum_{n> \nobs}\frac{c_{n}}{n!}\xi^n\ , 
\ee
which is passed on to the observational predictions. In particular, the spectral index and tensor-to-scalar ratio are expected to be off by an amount $\Delta n_s$ and $\Delta r$ respectively, where 
\be
\frac{|\Delta n_s|}{1- n_s} \sim \frac{|\Delta r|}{ r} \sim \frac{\Delta \mathcal{C}_{{\rm eff}}}{ \mathcal{C}_{{\rm eff}}}
\ee
The key point is that the data itself contains errors in the values of these parameters. Of relevance to us,  the combined measurement of \textit{Planck} TT,TE,EE+lowE+lensing+BK15+BAO \cite{1807.06211}  yields
 \be
n_s=0.9668\pm0.0037, \qquad  r < 0.063\ ,
\ee
where the confidence interval is to $1\sigma$ and the upper bound on $r$ to $2\sigma$. The errors introduced through our truncation will be less than the observational errors in the data provided  $\Delta \mathcal{C}_{{\rm eff}}/ \mathcal{C}_{{\rm eff}} \lesssim 0.1$ for the generic potentials (and their derivatives).  In this work we truncate the potentials at 20th order, for which $\nobs=20$. To get a flavour of the size of  $\Delta \mathcal{C}_{{\rm eff}}/ \mathcal{C}_{{\rm eff}}$, consider the case where $\mathcal{C}_{{\rm eff}}=e^\xi$.  In that case, the generic error can be expressed in terms of incomplete Gamma functions, 
\be
\frac{\Delta \mathcal{C}_{{\rm eff}}}{ \mathcal{C}_{{\rm eff}}}=1-\frac{\Gamma(\nobs+1, \xi)}{\nobs!}
\ee
For $\nobs=20$ and $\xi$ bounded by the strong coupling limit,  $\xi \leq 4\pi$, we find that $\Delta \mathcal{C}_{{\rm eff}}/ \mathcal{C}_{{\rm eff}} \le 1-\frac{\Gamma(21, 4\pi)}{20!} 
\approx 0.018$. This places the error well below the observational limits. 

For the truncated polynomials, the slow-roll equations were solved numerically using \textsc{Python}. For a given model, the procedure we adopt is as follows:
\begin{enumerate}
    \item Initially, all roots $\mathcal{V}_{\text{eff}} (\xi) =0$ in the range $\xi \in [0.01, 4\pi]$ are identified, and any intervals with $\mathcal{V}_{\text{eff}}<0$ are discarded. The lower bound in $\xi$ arises from noting that for $\xi<1$ the potential becomes nearly quadratic which would be in tension with data. As a sanity check, we allow a region $0.01\leq\xi\leq1$ where we expect quadratic inflation to occur and show that observationally viable inflation requires $\xi>1$.
    \item Within each interval, $\xi\in[\xi_{\rm lower}, \xi_{\rm upper}]$, we solve for all possible field values that will lead to the end of inflation, specified by $\epsilon (\xi_e) = 1$. 
    \item For each $\xi_e$, we evaluate whether $N_{\star}$ can be satisfied using~(\ref{eqn:efolds}). This is achieved by choosing a trial value for $\xi_{\star} = \xi_{e} + \Delta \xi$, and increasing $\Delta \xi$ until sufficient e-folds are achieved, or $\xi_{\star} > \xi_{\rm upper}$. If successful, a further root finding step is performed to find  $\epsilon_{\star}$. 
    \item Finally, we perform an additional check that both slow-roll conditions are satisfied, $|\eta| < 1, \epsilon < 1 \, \forall \xi\in\left[\xi_e,\xi_{\star}\right]$. 
\end{enumerate}
Passing these conditions ensures there is a sufficient period of slow-roll inflation. Each model may give rise to either no, single or multiple inflationary regions. Before discussing the generic predictions for these families of models, let us point out a few important caveats as well as fleshing out the numerical strategy that we followed. In the following, we will only consider models that comply with the following:
\begin{enumerate}[leftmargin=3\parindent, label=\roman*.]
\item Positivity of $\mathcal{Z}_{\text{eff}}$: In order to avoid ghosts in the theory, we require that 
\begin{equation*}
\mathcal{Z}_{\text{eff}}(\xi)>0\ \forall \xi\in\left[\xi_e,\xi_{\star}\right]\ .
\end{equation*}
\item After the end of inflation, the potential monotonically decreases to the Minkowski vacuum defined at $\xi=0$. The original Kaloper-Sorbo theory is defined as an expansion around the Minkowski vacuum at $\xi=0$, the weakly coupled regime. We limit ourselves to those potentials that allow access to this vacuum classically. 
\item Related to the previous point, we only consider the classical evolution of the field. Even though one might be tempted to argue in favour of keeping models with false de Sitter vacua, since in some finite time we would tunnel to the true Minkowski vacuum, the technicalities of the calculation involving tunnelling probabilities disallow the statistical treatment of the next section.
\item We only consider a single phase of inflation. Although models with multiple phases of inflation could satisfy the observational constraints on the length of inflation, these models require further study of the dynamics and the spectrum of perturbations. This implies that, in the case of multiple regions that satisfy the above constraints, we choose the region that is closest to the Minkowski vacuum, as this would be the first N e-folds of inflation that we would observe.
\end{enumerate}

\begin{figure}[ht!]
	\centering
	\includegraphics[width=0.7\textwidth]{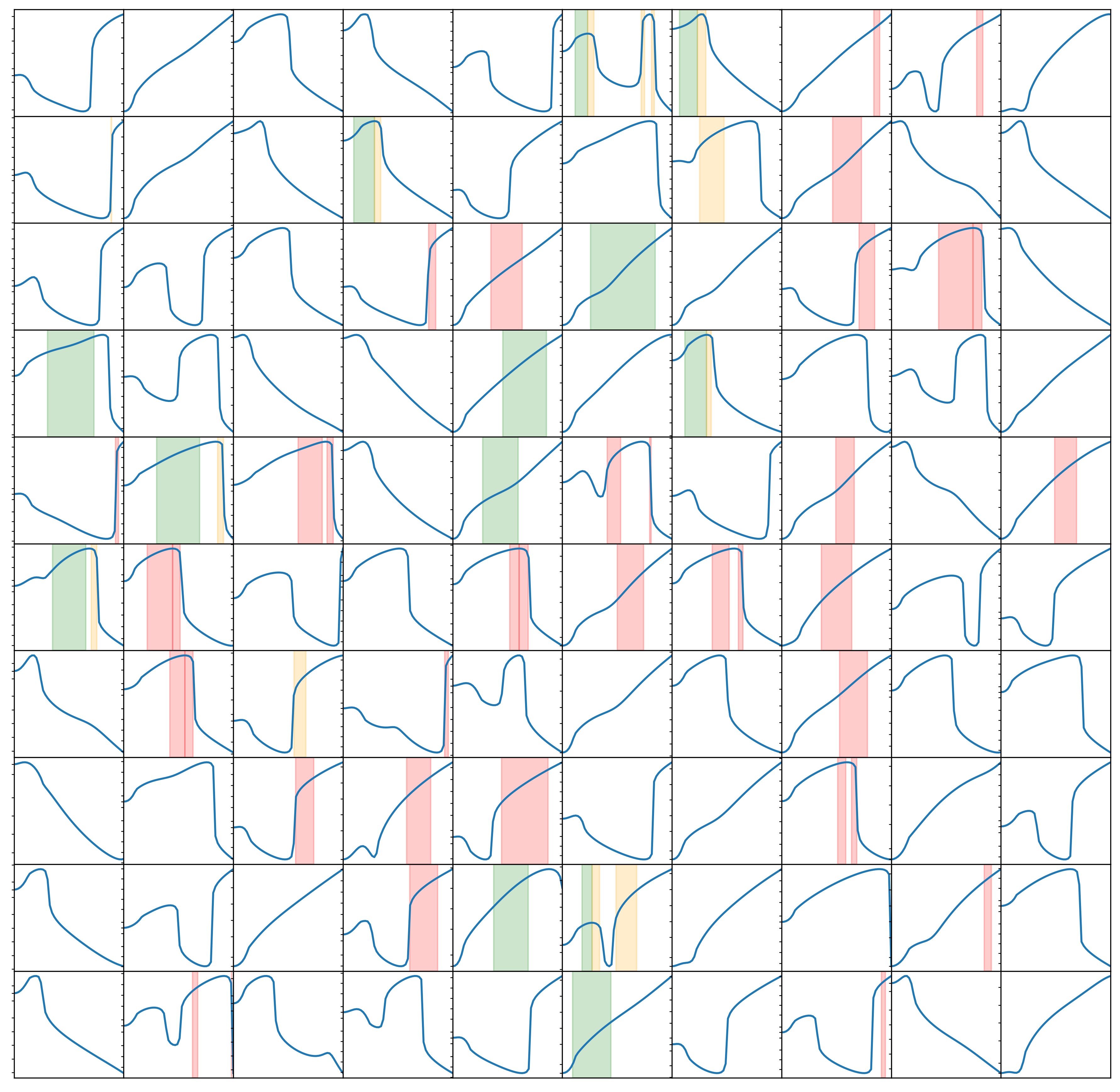}
	\caption{Random potentials, $\mathcal{V}_{\text{eff}}$ v $\xi$, for the Kaloper-Lawrence model. The banded regions show $\left[\xi_e,\xi_{\star}\right]$. The red regions are ruled out by the positivity of $\mathcal{Z}_{\text{eff}}$ and the orange regions by the requirement that the final phase of evolution must  decrease monotonically to the Minkowski vacuum.} 
	\label{random_models}
\end{figure}
 In order to gain some intuition on the range of possible solutions, we generate 10k Kaloper-Lawrence models drawn randomly from a mass prior of $m \sim \mathcal{U} (0.1,1)$, and $b_{n>2}$, $c_{n>0} \sim \mathcal{U}\left[-3,-0.1\right]\cup \mathcal{U}\left[0.1,3\right]$. The remaining parameters are set by their values in \eqref{eq:L_EFT}, {\it i.e.} $b_{0},b_1=0$, $b_2=1$ and $c_{0}=1$. These priors are chosen to give natural $\mathcal{O}(1)$ values. We also did a run with coefficients in the range $\mathcal{U}\left[-3,3\right]$ with no major change in results. During sampling, we choose $N_{\star} \sim \mathcal{U} (50,60)$ to give a representative range consistent with standard reheating. As discussed above, we truncate the series at $\nobs=20$.
 
\begin{figure}[ht!]
	\centering
	\includegraphics[width=\textwidth]{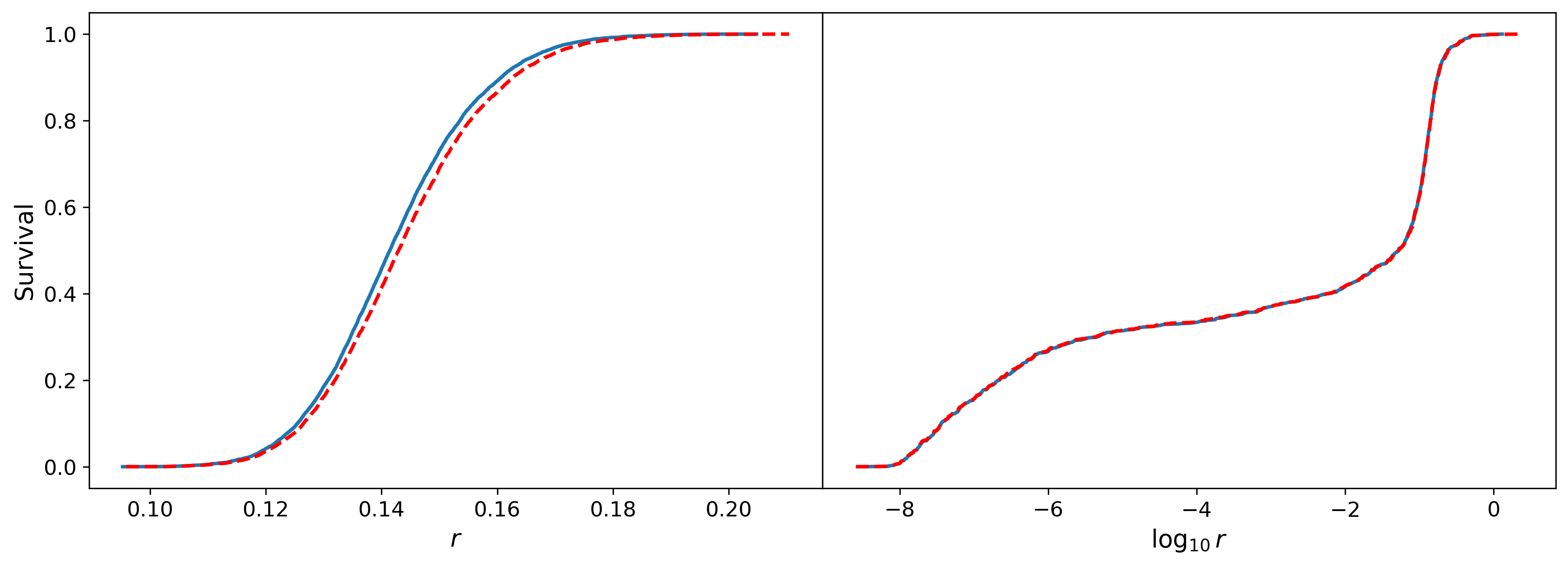}
	\caption{Fraction of viable inflationary models that survive an observational bound on $r$. On the left we show positive coefficient models only, and on the right both positive and negative coefficients. The same model but with VES included is given by the dashed red curve.} 
	\label{survival}
\end{figure}

In total, we find 39\% of models satisfy slow-roll and give a sufficient number of e-folds, but only 12\% are viable, due to conditions (i) and (ii) above\footnote{We do not consider observations when determining if a model is viable.}. We plot a selection of models in Fig.~\ref{random_models}. If one repeats the analysis considering {\em only} positive coefficients, we find viable inflation occurs for all 10k random models. Many of these models, of course, will not give consistent observations, so in Fig.~\ref{survival} we show the fraction of viable models that survive a given observational bound on $r$. Of particular note is that for KL models both with and without VES enhancement, if we demand only positive coefficients, none of them in the random sample survive $r < 0.1$, whereas, when we allow for both positive and negative coefficients, $\sim 30$\% of the previously viable models survive the very stringent CMB bound of $r < 0.001$. 

To evaluate the impact of VES, we use priors of $\sb_{n>1},  \sc_{n>1} \sim \mathcal{U}\left[-3,-0.1\right]\cup \mathcal{U}\left[0.1,3\right]$ and $\log_{10} \kappa \sim \mathcal{U} (-5,-1)$. The ranges on $\kappa$ are fixed so that the condition $\kappa:=\bar m / m\ll1$ is satisfied, with the upper bound marginally doing so. One could consider even lower values of $\kappa$, however that would make the contributions from the VES sector even  weaker. For the range $10^{-5}<\kappa<10^{-1}$, it is already the case that including VES does not significantly change these results and one finds a similar number of viable models. Therefore, we would expect that pushing the bound on $\kappa$ to lower values would not yield any interesting new  results. 

The small impact arising from the VES corrections are not entirely unexpected. For example, from \eqref{eq:d_V_eff} and using that the sum peaks around $\xi\simeq n+1$ we see that the first derivative of the potential is given by
\be
\mathcal{V}'_{\text{eff}}= {m\over \mu^2}\sum_{n\geq2}\left(n\,b_n + \sb_{n,1}\kappa\, \xi\right)\frac{\xi^{n-1}}{n!} \sim  {m\over \mu^2}\sum_{n\geq2} n(1+\kappa)\frac{\xi^{n-1}}{n!}\sim \mathcal{V}'_{\text{eff}}(\xi,0)(1+\kappa)\ ,
\ee
with the corrections to the cosmological observables being, roughly, $n_s \sim n_s^{\rm ns}(1+\kappa)$ and $r \sim r^{\rm ns}(1+\kappa)$, with $n_s^{\rm ns}, r^{\rm ns}$ the original non-VES KL predictions. Nonetheless, it is very interesting to point out that despite more than doubling the number of parameters in the theory, the VES models are not disfavoured by a Bayesian analysis, as shown in \secref{sec:MCMC}. From a theoretical  point of view, the VES models might even be preferred since they include a mechanism to cancel vacuum energy loop contributions. 

 In the following section we assess more concretely whether these solutions give rise to slow-roll parameters consistent with observations.

\section{Testing with observations and forecasting}
\label{sec:MCMC}
We now ask whether the models under investigation   give rise to slow-roll parameters consistent with observations.
We first consider the case of strictly positive Wilson coefficients using \textit{Planck} 2018~\cite{Planck:2018jri} and Bicep-Keck (BK15)~\cite{Ade:2018gkx} data. As suggested by Fig.~\ref{survival}, we expect these class of models to be under tension due to the paucity of consistent solutions satisfying $r < 0.1$. However, these were generated by random sampling from the prior, and it may be possible that regions of parameter space are more consistent with observations. 

\begin{figure}[ht!]
\centering
\includegraphics[width=1.0\textwidth]{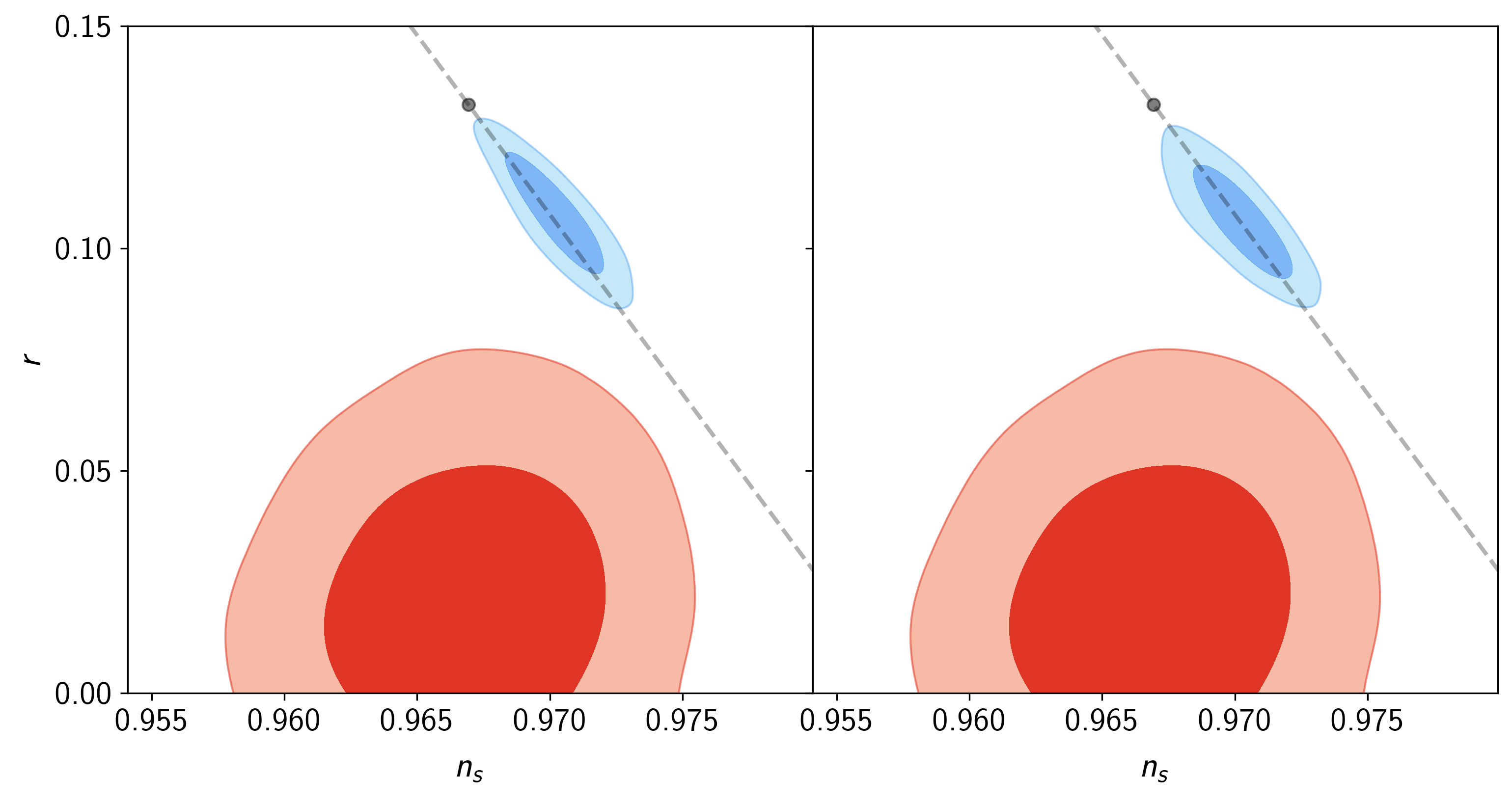}
\caption{\textit{Planck} data (red) along with predictions (blue) for Kaloper-Lawrence inflation with strictly positive Wilson coefficients, both without (left) and with (right) VES contributions. We also show the $V\sim \phi^p$ monodromy line, with $p=2$ and $N_{\star}=60$ given by the filled circle.}
\label{fig:planck}
\end{figure}

For this analysis, we perform a Markov Chain Monte Carlo (MCMC) analysis using a Gaussian likelihood. We fit the ($n_\mathrm{s}$, $r$) covariance matrix from the  \textit{Planck} TT, TE, EE + lowE + BAO + BK15 chains, which gives a good approximation to the true likelihood. In total we fit for 39 parameters in the Kaloper-Lawrence model, and 79 including VES enhancements, using the same prior ranges as in the previous section and fixing $N_{\star}=60$. We use the ensemble sampler \textsc{emcee}~\cite{2013PASP..125..306F}, using 200 walkers in the ensemble and a combination of the affine invariant stretch~\cite{2010CAMCS...5...65G} and differential evolution moves. The reason for this is when generalising to negative Wilson coefficients we find a bi-modal posterior, with a mode of lower likelihood running along the monodromy line, and this combination of MCMC moves gives improved mixing. Running the ensemble for a chain length of 20 autocorrelation times, the resulting posterior is shown in Fig.~\ref{fig:planck}. As expected, models with strictly positive Wilson coefficients are unable to fit  observational data and can be ruled out at the $\sim 2.7\sigma$ level.

The marginalised parameters and best-fits are similar both with and without VES, as expected from the small $\sim (1+\kappa)$ corrections to $r$ and $n_\mathrm{s}$. This is supported by inspecting the posterior distribution for $\kappa$, which only becomes non-uniform for $\log_{10} \kappa \gtrsim -2$. Given the large number of additional parameters in the fit, it is natural to ask if VES is disfavoured by model comparison. In order to quantity this, we compute the Bayesian Evidence by performing nested sampling. For ease of integration with our code we use the \textsc{dynesty} sampler~\cite{Speagle:2019ivv} with 1000 live points. The Bayesian Evidence ($B$) values are given in \tabref{table:parameters}. We find the difference between VES and no VES to be $\log B=0.6$, which is inconclusive on the Jeffreys scale. Whilst this may seem counter-intuitive, VES is compatible with data over almost all of its prior range, so is not penalised by Bayesian Evidence as very little additional parameter space is wasted.

\def\arraystretch{1.0} %vertical space
\begin{table}
%\capstart
\centering
\resizebox{\textwidth}{!}{\begin{tabular} {| c | c | c | } \hline
 Dataset & Model &  $\log B$  \\ \hline\hline
\textit{Planck} & Positive Wilson coefficients  & -4.8 \\ \hline
\textit{Planck} & Positive Wilson coefficients with VES & -4.2 \\ \hline
\textit{Planck} & Positive and negative Wilson coefficients  & -4.4 \\ \hline
\textit{Planck} & Positive and negative Wilson coefficients with VES & -3.8 \\ \hline \hline
CMB S3 & Positive Wilson coefficients  & -40.5 \\ \hline
CMB S3 & Positive Wilson coefficients with VES & -41.0 \\ \hline
CMB S3 & Positive and negative Wilson coefficients  & -26.9 \\ \hline
CMB S3 & Positive and negative Wilson coefficients with VES & -24.3 \\ \hline  \hline
CMB S4 & Positive Wilson coefficients  &  -231.7 \\ \hline
CMB S4 & Positive Wilson coefficients with VES &  -229.5 \\ \hline
CMB S4 & Positive and negative Wilson coefficients  & -27.2 \\ \hline
CMB S4 & Positive and negative Wilson coefficients with VES & -28.0 \\ \hline  \hline
\end{tabular}}
\caption{Log of the Bayesian Evidence $B$ for various data and model combinations. }
\label{table:parameters}
\end{table}

\begin{figure}[ht!]
\centering
\includegraphics[width=1.0\textwidth]{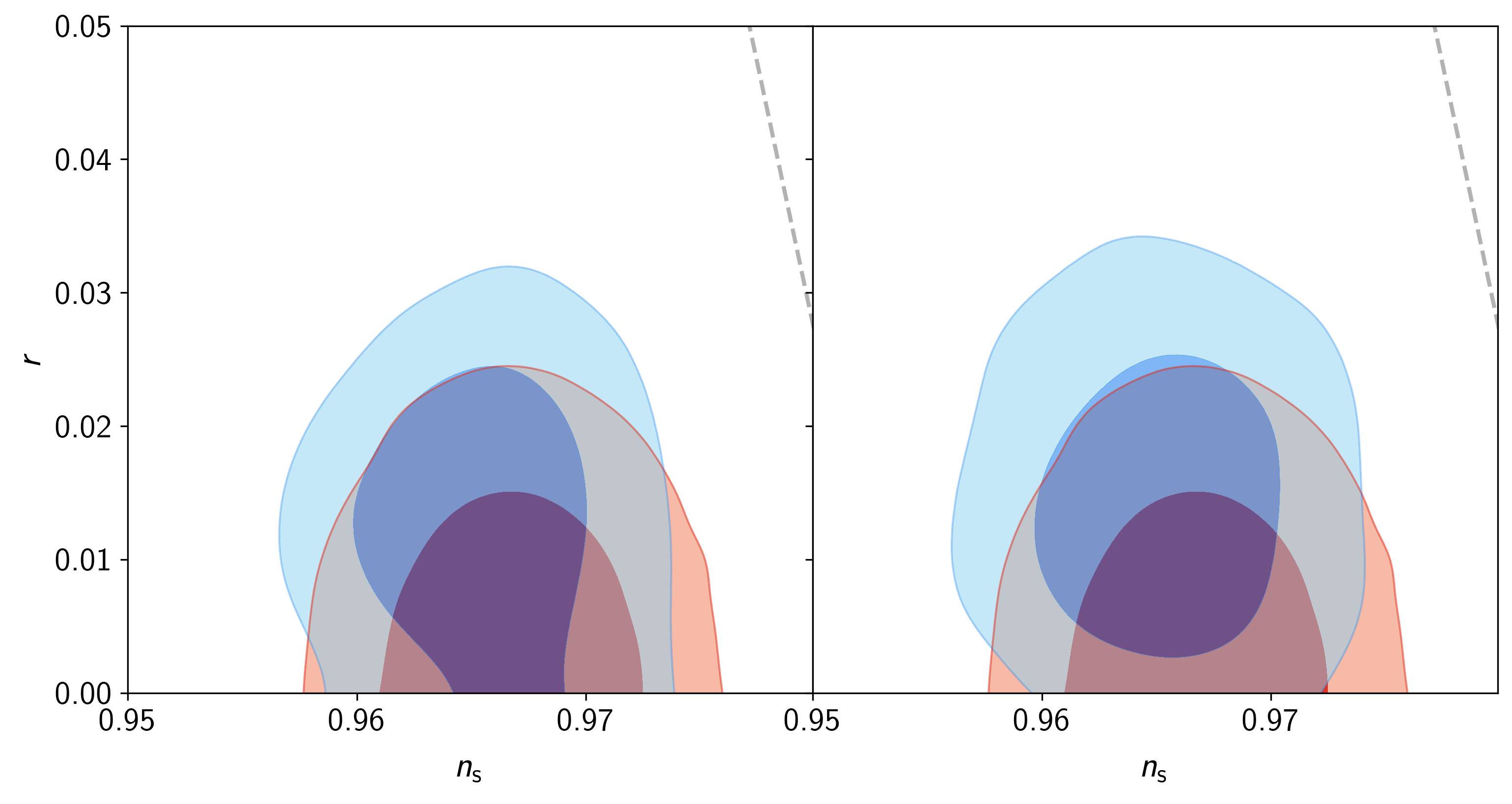}
\caption{ CMB S3 experimental data (red) along with predictions (blue) for Kaloper-Lawrence inflation with positive and negative Wilson coefficients, both without (left) and with (right) VES contributions.}
\label{fig:cmbs3}
\end{figure}

\begin{figure}[ht!]
\centering
\includegraphics[width=1.0\textwidth]{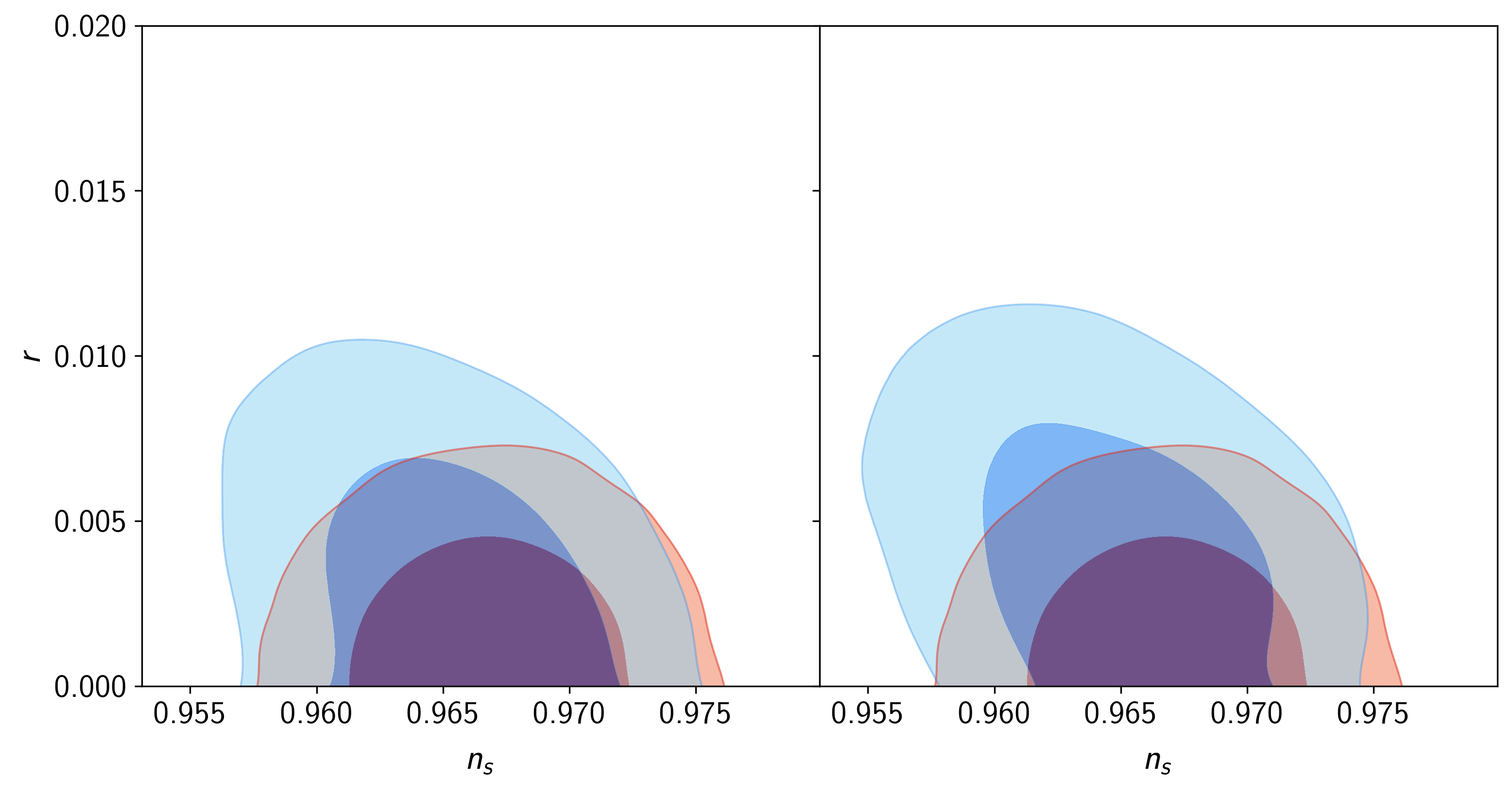}
\caption{CMB S4 experimental data (red) along with predictions (blue) for Kaloper-Lawrence inflation with positive and negative Wilson coefficients, both without (left) and with (right) VES contributions.}
\label{fig:cmbs4}
\end{figure}

The result of Fig.~\ref{survival} indicates that negative Wilson coefficients are needed. This is related to the fact that observationally successful models of inflation require the presence of a turning point in the potential, {\it i.e.} some degree of cancellation between neighbouring coefficients. Running the MCMC, we find this class of model has no issue in fitting current data, so we also check that it is compatible with forecasted Stage 3 (S3)  and  Stage 4 (S4) constraints. To do this, we assume an uncorrelated Gaussian  likelihood for ($n_\mathrm{s}$, $r$), using $1\sigma$ upper bounds on $r$ of 0.01 and 0.003 respectively. Since these experiments will primarily target the large-angle polarization signal, we use existing \textit{Planck} constraints on $n_\mathrm{s}$. In Figs.~\ref{fig:cmbs3} and~\ref{fig:cmbs4} we show the resulting posterior distributions. As expected from the previous section, these models can naturally give rise to small $r$ values and cannot be ruled out, even by a future S4 experiment. 

\begin{figure}[ht!]
\centering
\includegraphics[width=1.0\textwidth]{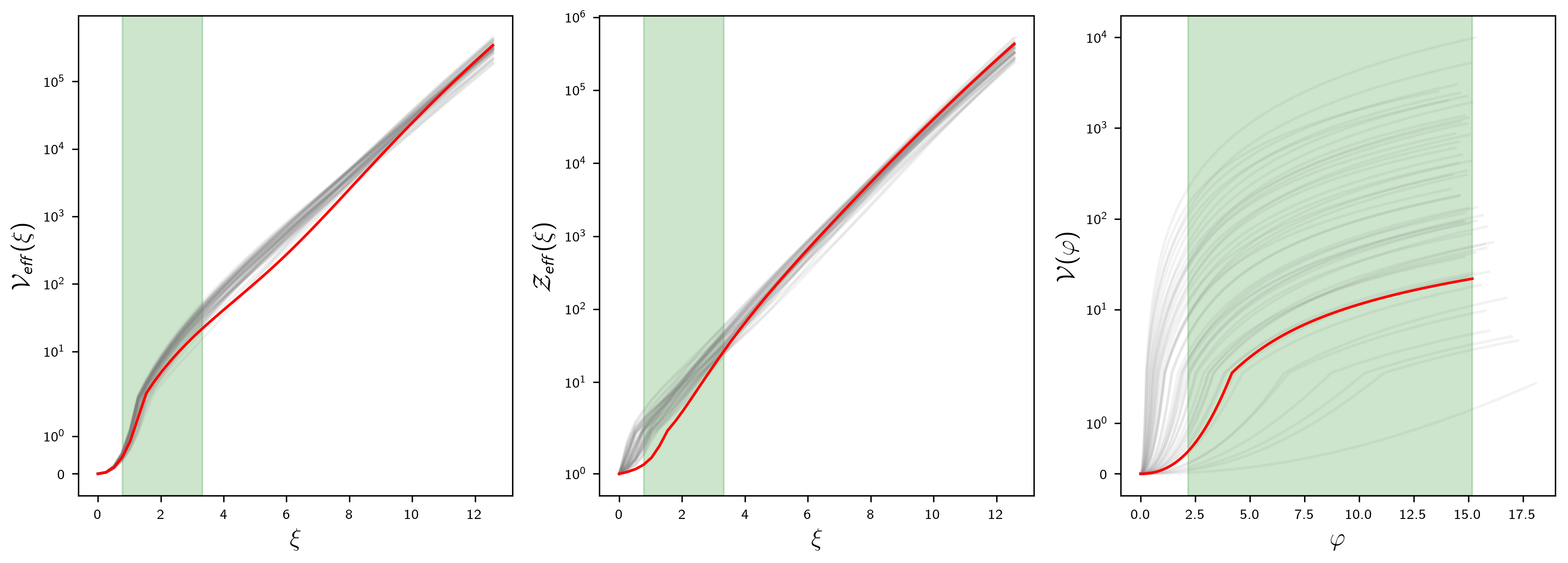}
\caption{ (Left) $\mathcal{V}_{\text{eff}}(\xi)$ from MCMC samples for the Kaloper-Lawrence model with positive Wilson coefficients and \textit{Planck} data. (Middle) $\mathcal{Z}_{\text{eff}}(\xi)$. (Right) Potential for the canonical field,  $\mathcal{V}_{\text{eff}}(\psi)$. The banded regions show the final 60 e-folds of inflation. }
\label{fig:potential_reconstruction}
\end{figure}

\begin{figure}[ht!]
\centering
\includegraphics[width=1.0\textwidth]{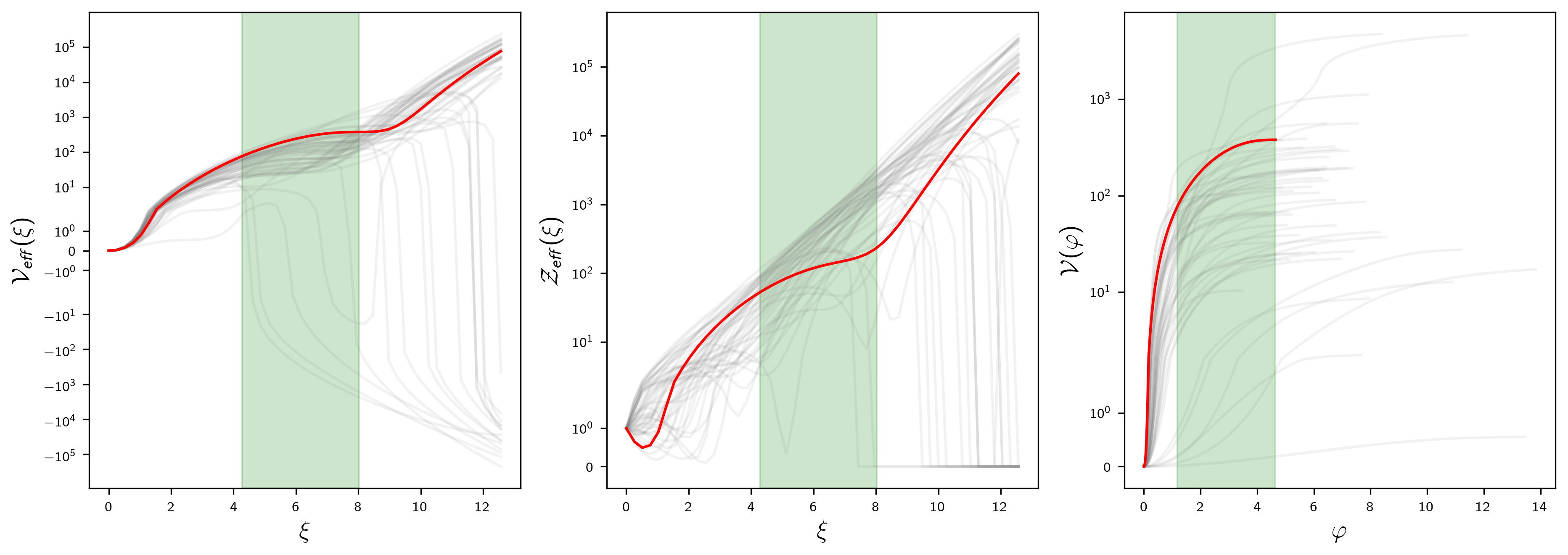}
\caption{Same as \figref{fig:potential_reconstruction} but for positive and negative Wilson coefficients with S4 constraints. The combination of positive and negative coefficients flattens the potential during the crucial final 60 efoldings of inflation region, leading naturally to small values of $r$ consistent with the data. }
\label{fig:potential_reconstruction2}
\end{figure}

To see why, it is worth examining the reconstruction of the potential. In Fig.~\ref{fig:potential_reconstruction} we show $\mathcal{V}_{\text{eff}}(\xi)$, $\mathcal{Z}_{\text{eff}}(\xi)$ and $\mathcal{V}_{\text{eff}}(\varphi)$ for samples from the MCMC chains for the Kaloper-Lawrence model with positive Wilson coefficients and \textit{Planck} data. In Fig.~\ref{fig:potential_reconstruction2} we show the same but for positive and negative coefficients and a S4 experiment. The crucial role played by the negative Wilson coefficients becomes clear -- they have the effect of flattening the potential from its otherwise rapdily monotonic decline, in particular leading to  a point of inflection from where inflation can begin, staying flat enough to give rise to at least sixty efolds of acceleration expansion. This flatness slows the inflaton down, decreasing $\epsilon$, and leading to a much smaller value for $r$.

Plotting a histogram of the number of negative coefficients, we find this is strongly peaked at $\sim n / 2$. The requirement that a degree of cancellation is required between coefficients is not ideal from a model building perspective.  This can be quantified by comparing the Bayesian Evidence between models with positive and negative coefficients, compared to strictly positive coefficients. From \tabref{table:parameters} one can see that for \textit{Planck}, even though there is an improvement in the fit, there is only weak evidence for including negative coefficients. For a stage 3 experiment, however, if there is no detection of tensor modes, there would be strong evidence for negative coefficients in the context of monodromy models. 

\section{Conclusions \label{sec:conc}}

Field theory models of monodromy inflation allow one to consistently probe large values of the inflaton field whilst retaining control of the effective description.  Of course, such models are guaranteed to be ruled out by observation unless there is some additional flattening of the potential, something which can be achieved by pushing the dynamics into a strongly coupled regime,  without losing control of the theory. Although an heuristic analysis of the observational tests of these models was performed in \cite{1709.07014}, a thorough numerical examination was still lacking. In this paper, we have carried out that examination and revealed some new features required by the family of models in order for them to satisfy the observational constraints.  In particular, the tower of Wilson coefficients that define the higher order operators of the theory cannot all have the same sign. Indeed,  in order for the potential to be sufficiently flat to yield enough inflation and be compatible with bounds on the spectral index and tensor to scalar ratio, there needs to be cancellations between terms. This requires a combination of positive and negative coefficients. Once this is the case, these coefficients can be order one in absolute value,  consistent with naturalness,  and remain compatible with CMB data, both current and forecasted Stage 3 and Stage 4 with strong constraints on the tensor to scalar ratio. Although the best models seem to require a roughly equal number of positive and negative coefficients, we were not able to identify any further structure in the array of coefficients.

These monodromy models can be enhanced, from a theoretical perspective by including the corrections suggested by vacuum energy sequestering \citeseq. These are desirable since they guarantee the cancellation of vacuum energy loops, allowing  us to set the vev of the potential to vanish without having to fine tune the result against large radiative corrections. From a Bayesian perspective,  we might have expected the addition of the VES corrections to have been disfavoured because we introduce many more parameters into the theory. As it happens, this doesn't matter:  VES enhanced models are not significantly favoured or disfavoured relative to the original models of monodromy inflation, at least as far as the data is concerned.  This is encouraging as it suggests an emergent  mechanism for solving the cosmological constant problem can be incorporated into inflationary dynamics without any cost to the likelihood. 

We might ask whether there is any prospect for picking out VES enhanced models versus say the KL models? These initial results suggest not at the level of the inflaton itself, as the two models agree so closely. However there may be features arising during the reheating phase, arising from the extra couplings associated with VES models that would lead to observational features on small scales. We might also ask, what are the weaknesses and strengths of the scenarios we have investigated. A clear strength of the VES enhanced models is that they contain a mechanism to address the cosmological constant problem. A potential strength of both sets of models, with and without VES enhancement, arises from the fact that when allowing for positive and negative coefficients, these can conspire to flatten the potential as seen in Fig.~\ref{fig:potential_reconstruction}. It opens up the intriguing possibility that they may cause the potential to flatten to such a degree that the inflaton field could enter a period of ultra slow roll inflation, leading to the associated production of primordial black holes, and therefore providing, both, a new observational signature for the model and a new set of constraints on the allowed values of the coefficients. It would also arguably be the first particle inspired model that leads naturally to a period of ultra slow roll inflation occurring during the final e-foldings of inflation. Perhaps a weakness in probing both sets of models is that we don't know precise values of the Wilson coefficients and so the best we can do is to sample from a broad class of possible values. However, we are not alone with this problem, it is true of any effective theory unless there is some additional symmetry controlling the relative structure of the higher order operators.  Not knowing these coefficients makes it hard to say much about the end of inflation if you want to think about the formation of oscillons, axion stars, and possible links to dark matter. But it remains the case that we may be able to make more general statements about the likely formation mechanisms and the associated production of primordial gravitational waves - an exciting prospect.

\acknowledgments

AP, EJC and AM were supported by an STFC consolidated grant number ST/T000732/1 and FC by a University of Nottingham studentship. EJC is also supported by a Leverhulme Research Fellowship reference: RF-2021-312. 
AM is also supported by a Royal Society University Research Fellowship.
\newline

\noindent For the purpose of open access, the authors have applied a CC BY public copyright licence to any Author
Accepted Manuscript version arising.
\newline

\noindent Data Availability Statement: The code and data produced in this paper are available upon reasonable request to AM.

%This work is entirely theoretical and has no associated data.

\bibliography{bib}

\end{document}